\documentclass[aps, prb, floatfix, reprint, superscriptaddress, showpacs]{revtex4-1}

\usepackage{graphicx,amsmath,amssymb,bm}

\begin{document}

\title{Contribution of the magnetic resonance to the third harmonic generation from a fishnet metamaterial} 

\author{J. Reinhold}
\email{reinhold@uni-jena.de}
\affiliation{Institute of Applied Physics, Abbe Center of Photonics, Friedrich-Schiller-Universit\"at Jena, Max-Wien-Platz 1, 07743 Jena, Germany}

\author{M. R. Shcherbakov}
\email{shcherbakov@nanolab.phys.msu.ru}
\affiliation{Faculty of Physics, Lomonosov Moscow State University, Moscow 119991, Russia}

\author{A. Chipouline}
\affiliation{Institute of Applied Physics, Abbe Center of Photonics, Friedrich-Schiller-Universit\"at Jena, Max-Wien-Platz 1, 07743 Jena, Germany}

\author{V.~I.~Panov}
\affiliation{Faculty of Physics, Lomonosov Moscow State University, Moscow 119991, Russia}

\author{C. Helgert}
\affiliation{Institute of Applied Physics, Abbe Center of Photonics, Friedrich-Schiller-Universit\"at Jena, Max-Wien-Platz 1, 07743 Jena, Germany}

\author{T.~Paul}
\affiliation{Institute of Condensed Matter Theory and Solid State Optics, Abbe Center of Photonics, Friedrich-Schiller-Universit\"at Jena, Max-Wien-Platz 1, 07743 Jena, Germany}

\author{C.~Rockstuhl}
\affiliation{Institute of Condensed Matter Theory and Solid State Optics, Abbe Center of Photonics, Friedrich-Schiller-Universit\"at Jena, Max-Wien-Platz 1, 07743 Jena, Germany}

\author{F. Lederer}
\affiliation{Institute of Condensed Matter Theory and Solid State Optics, Abbe Center of Photonics, Friedrich-Schiller-Universit\"at Jena, Max-Wien-Platz 1, 07743 Jena, Germany}

\author{E.-B.~Kley}
\affiliation{Institute of Applied Physics, Abbe Center of Photonics, Friedrich-Schiller-Universit\"at Jena, Max-Wien-Platz 1, 07743 Jena, Germany}

\author{A. T\"unnermann}
\affiliation{Institute of Applied Physics, Abbe Center of Photonics, Friedrich-Schiller-Universit\"at Jena, Max-Wien-Platz 1, 07743 Jena, Germany}

\author{A. A. Fedyanin}
\affiliation{Faculty of Physics, Lomonosov Moscow State University, Moscow 119991, Russia}

\author{T. Pertsch}
\affiliation{Institute of Applied Physics, Abbe Center of Photonics, Friedrich-Schiller-Universit\"at Jena, Max-Wien-Platz 1, 07743 Jena, Germany}

\date{\today}

\begin{abstract}
We investigate experimentally and theoretically the third harmonic generated by a double-layer fishnet metamaterial. To unambiguously disclose most notably the influence of the magnetic resonance, the generated third harmonic was measured as a function of the angle of incidence. It is shown experimentally and numerically that when the magnetic resonance is excited by a pump beam, the angular dependence of the third harmonic signal has a local maximum at an incidence angle of $\theta\simeq20^\circ$. This maximum is shown to be a fingerprint of the antisymmetric distribution of currents in the gold layers. An analytical model based on the nonlinear dynamics of the electrons inside the gold shows excellent agreement with experimental and numerical results. This clearly indicates the difference in the third harmonic angular pattern at electric and magnetic resonances of the metamaterial.
\end{abstract}


\maketitle

\section{Introduction}
The emergent field of meta\-materials has brought optical materials to a qualitatively new level. It became possible to access new functionalities and optical properties of media by applying subwavelength structuring. Spectral selectivity and extraordinary optical transmission, \cite{Ebbesen1998} chirality,\cite{Decker2007} anisotropy,\cite{Bryan-Brown1990, Shcherbakov2010} optical magnetism, and negative refraction could be assigned to thin-film media by means of modern nanolithography.\cite{Soukoulis2007,Zhang2005,Dolling2007,Shalaev2005,Enkrich2005} The majority of effects in optical metamaterials arise due to the excitation of plasmon polaritons at interfaces between metal inclusions and surrounding dielectrics. Plasmon polaritons produce highly localized electromagnetic field densities, making metamaterials attractive from the point of view of nonlinear optical effect enhancement \cite{Kim2008,Klein2006,Niesler2011,Wuestner2011} and tailoring of nonlinear optical properties.\cite{Kujala2007,Kujala2008,Husu2012,Gentile2011,Utikal2010,Utikal2011,Lapine2012,Litchinitser2009,Poutrina2010} Magnetic metamaterials, i.e., metamaterials that mimic optical magnetism by supporting circular current plasmonic modes,\cite{Podolsky2002,Enkrich2005} are of special interest since the respective circular currents can play a significant role in the nonlinear optical response,\cite{Klein2006,Klein2007,Zeng2009,Tang2011} and considerable effort was directed towards determining the peculiarities of the nonlinear optical response caused by the excitation of magnetic resonances.\cite{Klein2006,Klein2007,Kim2008} 

In this paper we attempt to demonstrate an implicit evidence of the symmetry-induced characteristics of the nonlinear response in magnetic metamaterials. Therefore we clearly disclose the magnetic mode contribution to the third-order nonlinearity of the fishnet metamaterial. This is done by means of angular spectroscopy of third harmonic generation (THG) and numerical modeling of THG with a Fourier modal method (FMM) with a nonlinear extension. The results are specific to the case when the magnetic resonance of the metamaterial is excited by pump radiation. The magnetic mode contribution arises from the antisymmetric current distribution in the two gold layers of the metamaterial and is revealed as a local maximum of THG intensity in the angular dependence at tilted incidence. The data is supported by an analytical model based on the dynamics of coupled nonlinear oscillators. This reveals the strong influence of the resonance symmetry on the third harmonic angular radiation pattern because of the retardation effects.

\section{Sample}

\begin{figure}[b]
\includegraphics[width=\columnwidth]{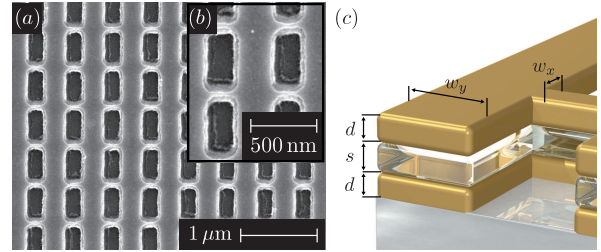}
\caption{(a) A scanning electron microscope (SEM) picture taken from the top view. The inset (b) shows a magnified SEM picture. The period $p$ is $500\,\text{nm}$ in both lateral directions. (c) The details of the geometry parameters are $w_x=110\,\text{nm}$, $w_y=290\,\text{nm}$, $d=23\,\text{nm}$, and $s=65\,\text{nm}$.}
\label{fig:sample}
\end{figure}

The fishnet structure is laterally defined by electron beam lithography (Vistec SB350OS) and a lift-off technique on a SiO$_2$ substrate and comprises a set of rectangular holes fabricated in a three-layer Au-MgO-Au heterostructure. The resulting structure has thin wires with a width of $w_x=110\,$nm and broad wires with a width of $w_y=290\,$nm. The structure has a period of $p=500\,$nm in both lateral directions. The thicknesses of both Au films are $d=23\,$nm and the thickness of the intermediate dielectric MgO film is $s=65\,$nm. Figures \ref{fig:sample}(a) and \ref{fig:sample}(b) show different scales of a scanning electron microscope (SEM) image of the sample taken from the top view. The parameters of the sample are shown in Fig.\ \ref{fig:sample}(c). The specific design parameters were chosen to match the magnetic resonance wavelength to the telecom wavelength range.

\section{Linear optical response}

\begin{figure}[b]
\includegraphics[width=\columnwidth]{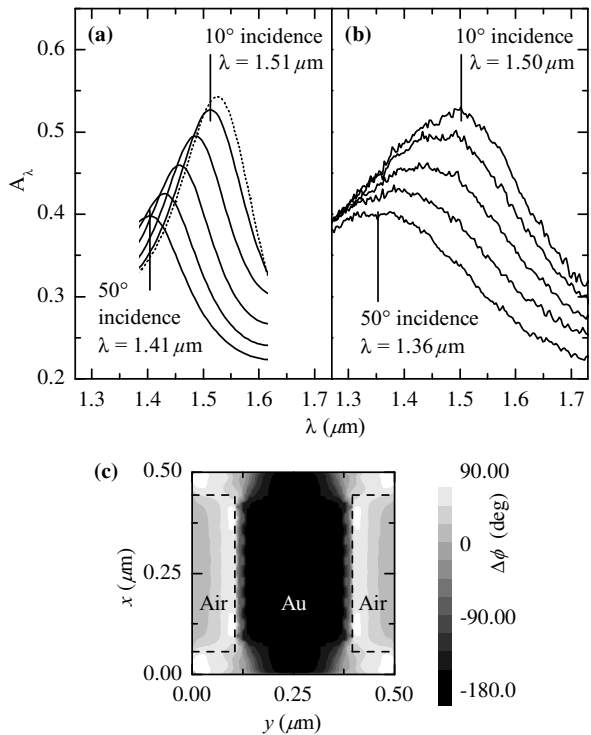}
\caption{Numerically simulated (a) and measured (b) linear absorption $A_\lambda$ as a function of wavelength $\lambda$ in the spectral range of the magnetic resonance, plotted for different angles of incidence: $\theta=10^\circ$, 20$^\circ$, 30$^\circ$, 40$^\circ$, and 50$^\circ$. The dotted line in the simulation plot shows the absorption for normal incidence. (c) The calculated phase difference $\Delta\phi$ between the $y$ components of the electric fields in the top and bottom gold layers at the center of the resonance peak for normal incidence, at 1.54$\,\mu$m.}
\label{fig:linear}
\end{figure}

Simulated and measured linear absorption spectra of the sample for different angles of incidence are shown in Figs.\ \ref{fig:linear}(a) and \ref{fig:linear}(b). The angles vary from 10$^\circ$ up to 50$^\circ$ with 10$^\circ$ steps. The linear absorption $A_\lambda$ is measured with an integrating sphere module of a Perkin Elmer Lambda 950 spectrometer in the spectral range of 1.20\,--\,1.80\,$\mu$m in 2\,nm steps. For the theoretical description of the problem we applied the FMM which allows solving the linear diffraction problem for an arbitrary anisotropic bi-periodic multilayer structure.\cite{Li2003} For the simulation we used the parameters of the sample as measured using the SEM. The angular spectroscopy of absorption measured with $p$-polarized incoming light is in agreement with the theoretical predictions, i.e.\ existence of an absorption peak at a wavelength of approximately $\lambda=1.54\,\mu$m for normal incidence which is blue-shifted as the angle of incidence is increased. The magnetic moment of the resonance results from the currents inside the broader wires of the metamaterial flowing in the opposite directions. This is demonstrated in Fig.\ \ref{fig:linear}(c) where the calculated phase difference $\Delta\phi$ between the electric field in the top and bottom gold layers is shown as a grayscale plot from the top view.

\section{Nonlinear optical response}

\begin{figure}[b]
\includegraphics[width=6.5cm]{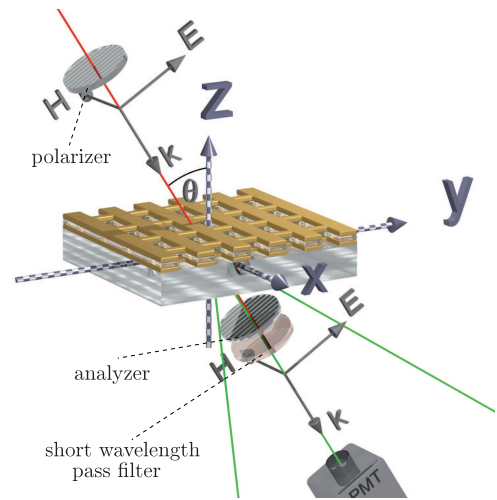}
\caption{The setup for angular spectroscopy of the third harmonic generation (THG) intensity. The pump polarization is set to $p$ and $p$-polarized third harmonic radiation is detected with a photomultiplier tube (PMT). The diffraction in the $x$ direction is not shown.}
\label{fig:setup}
\end{figure}

\begin{figure}
\includegraphics[width=\columnwidth]{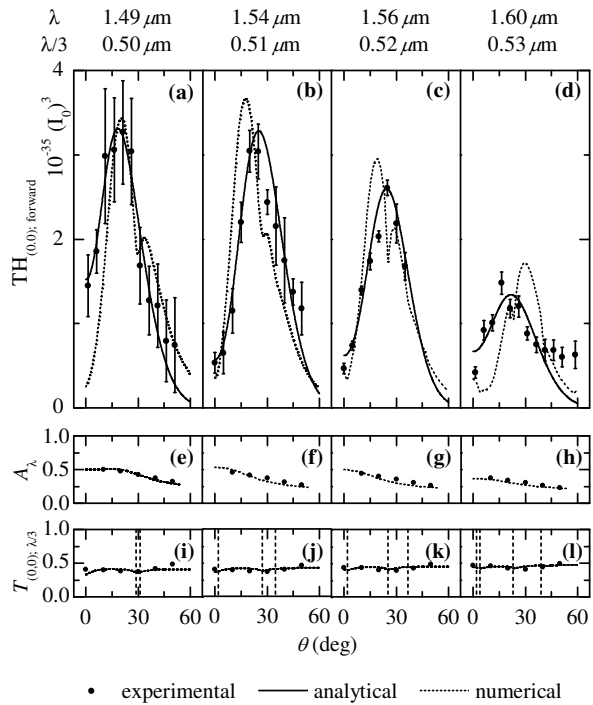}
\caption{(a)--(d) show the third harmonic signal as a function of the angle of incidence for different wavelengths in the spectral vicinity to the  magnetic resonance. For comparison, (e)--(h) show the linear absorption $A_\lambda$ at the same fundamental wavelengths and (i)--(l) show the linear transmission $T$ at the corresponding third harmonic wavelengths. The vertical dashed lines indicate the angular positions of the appearance and the disappearance of diffraction orders. The black dots represent the experimental data and the dotted lines represent the simulation results. The solid lines are curves calculated with Eq.\ (\ref{eq:fit}). This equation represents an analytical model which describes the nonlinear response of coupled oscillators.}
	\label{fig:thg}
\end{figure}

For the nonlinear measurements a setup based on an optical parametric amplifier (OPA) was used operating at wavelengths of 1.49, 1.54, 1.56 and 1.60$\,\mu$m and having an average output power of 3\,mW focused to a $300\,\mu$m spot from the air side of the sample. The OPA was pumped by a Nd:YAG laser with a pulse duration of 5\,ps and a repetition rate of 5\,kHz. The resulting fluence took values up to 700\,$\mu$J/cm$^2$ in the plane of the sample. The sample was placed on a six-axis positioning stage such that during the angular spectroscopy the beam is always focused into the same spot. The forward propagating THG signal pulses were detected by a photomultiplier tube and gate-integrated by an oscilloscope. We used the $p$-$p$ polarization configuration\,---\,illuminating with $p$-polarized light and selecting only the $p$-polarized part of forward propagating light before the detector. For all measurements spectral filtering (Schott RG610 and BG40) before the detector was used for picking up the desired wavelength. With these filters the third harmonic response was orders of magnitude larger than signals at other wavelength, i.e., at the pump wavelength. The averaged THG signal from the pure SiO$_2$ substrate measured outside the metamaterial area was at least one order of magnitude lower than that from the metamaterial area. Contributions from the substrate were therefore safely neglected. The principle setup is shown in Fig.\ \ref{fig:setup}.

For numerical simulation an extension of the FMM which includes the nonlinear interaction was used.\cite{Paul2010} The method relies on the undepleted pump approximation that ignores the feedback of the nonlinearity-induced field to the pump field.\cite{Boyd2003} The approach allows solving the problem completely rigorously and permits a reliable prediction of the diffracted amplitudes of the third harmonic fields.

The third harmonic intensity was measured and simulated in the forward zeroth diffraction order with the fundamental wavelength exciting the magnetic resonance. The angular spectra of THG are provided in Figs.\ \ref{fig:thg}(a)--\ref{fig:thg}(d) for the fundamental wavelengths of $1.49$, $1.54$, $1.56$, and $1.60\,\mu$m, respectively. The magnetic resonance position for normal incidence is $1.54\,\mu$m. The maximum of the THG signal is seen at angles of incidence around 20$^\circ$. The appearance of this maximum is detailed in the discussion section and is believed to be caused by the interference of THG from the individual layers forming the fishnet metamaterial. The simulation shows an agreement with the experimental values. The THG signal is expressed in a pump power-independent fashion as derived from the numerical calculations; the absolute values of the THG signal are valid only for the simulation results while for the experimental data they are of the same order of magnitude. The estimation of the experimental value of the effective nonlinear susceptibility is $\chi^{(3)}_{1111}=10^{-18}$\,m$^2$/V$^2$, which is the same order of magnitude as the reference value of bulk gold.\cite{Bloembergen1969} 

\section{Discussion}

Plasmon-enhanced THG at the magnetic resonance of fishnet metamaterials was reported previously.\cite{Kim2008} It was shown that the THG spectra obey the principles of the local-field enhanced nonlinear response. It was proposed that the wavelength dispersion of the THG efficiency is defined by the spectral line of the magnetic resonance cubed. The maximum of THG at angles of about 20$^\circ$ can neither be explained by means of dispersion of the local field factor at the fundamental frequency [see Figs.\ \ref{fig:thg}(e)--4(h)] nor with the linear transmission characteristics at the third harmonic wavelength [see Figs.\ \ref{fig:thg}(i)--4(l)]. Finally, the position of the maximum does not coincide with the angular position of the propagating diffraction order appearance as illustrated by the vertical dashed lines in Figs.\ \ref{fig:thg}(i)--4(l). In this section we show that, first, this feature is caused by retardation effects, and second, it is specific to the antisymmetric electric current structure of the magnetic resonance.

\begin{figure}
\begin{center}
\includegraphics[width=6.5cm]{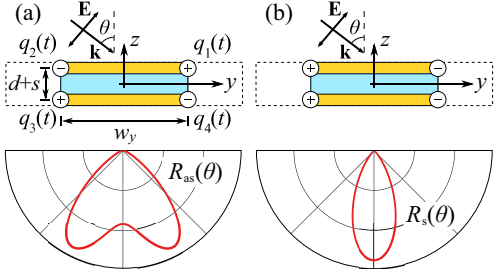}
\caption{\label{fig:fishnet_sketch} Parameters of the model and uncompensated charge density distribution in the unit cell of the fishnet metamaterial for (a) antisymmetric and (b) symmetric resonances and corresponding far-field radiation patterns. The blue area between the gold layers is shown for a better understanding of the layout, and no influence of the dielectric is assumed in the model.}
\end{center}
\end{figure}

The observed third harmonic radiation is considered to be caused by the nonlinear polarization of gold due to anharmonic electron movement. Nonlinearities of other substances of the metamaterial are neglected since their $\chi^{(3)}$-tensor components are several orders of magnitude smaller than that of bulk gold: $\chi^{(3)}_{1111}(\text{SiO}_2)=4.6\cdot10^{-23}\,\text{m}^2/\text{V}^2$, $\chi^{(3)}_{1111}(\text{MgO})=1.0\cdot10^{-22}\,\text{m}^2/\text{V}^2$, and $\chi^{(3)}_{1111}(\text{Au})=7.5\cdot10^{-19}\,\text{m}^2/\text{V}^2$.\cite{Adair1987,Adair1989,Bloembergen1969} Without further discussion on the specific source on that third-order nonlinearity, we describe the motion of electrons of gold at the third harmonic wavelength within the conducting layers of the metamaterial by using a model of weakly coupled oscillators. Within the chosen model the phase difference between the oscillators in the two layers dictates whether the resonance is antisymmetric\,---\,currents in the two layers are antiparallel to each other [Fig.\ \ref{fig:fishnet_sketch}(a)]\,---\,or symmetric\,---\,currents are parallel [Fig.\ \ref{fig:fishnet_sketch}(b)]. At the third harmonic wavelength this phase difference is assumed to be equal to the phase difference of the oscillators at a fundamental frequency multiplied by three. For the antisymmetric resonance the phase difference is equal to $\pi$, and for the symmetric one it is equal to zero (see the Appendix). 

With this knowledge we write down the dynamical equations for the charge density at the third harmonic frequency:
\begin{equation}
\begin{split}
\quad\rho^{as}({\mathbf r},t) &= q_0\cos 3\omega t\ \cdot \\
& \cdot\left[\delta\left(y-\frac{w_y}{2}\right)-\delta\left(y+\frac{w_y}{2}\right)\right] \cdot \\
& \cdot\left[\delta\left(z-\frac{d+s}2\right)-\delta\left(z+\frac{d+s}2\right)\right]
\end{split}
\end{equation}
and for the current density:
\begin{equation}\label{eq:jas}
\begin{split}
j_y^{as}({\mathbf r},t) &= 3\omega q_0\sin 3\omega t \cdot \\
& \cdot\left[\Theta\left(y-\frac{w_y}{2}\right)-\Theta\left(y+\frac{w_y}{2}\right)\right] \cdot \\
& \cdot\left[\delta\left(z-\frac{d+s}2\right)-\delta\left(z+\frac{d+s}2\right)\right]
\end{split}
\end{equation}
for the antisymmetric resonance and the dynamical equations for the charge density:
\begin{equation}
\begin{split}
\rho^{s}({\mathbf r},t)  &= q_0\cos 3\omega t\ \cdot \\
& \cdot\left[\delta\left(y-\frac{w_y}{2}\right)-\delta\left(y+\frac{w_y}{2}\right)\right] \cdot \\
& \cdot \left[\delta\left(z-\frac{d+s}2\right)+\delta\left(z+\frac{d+s}2\right)\right]
\end{split}
\end{equation}
and for the current density:
\begin{equation}\label{eq:js}
\begin{split}
j_y^{s}({\mathbf r},t) &= 3\omega q_0\sin 3\omega t\ \cdot \\
& \cdot\left[\Theta\left(y-\frac{w_y}{2}\right)-\Theta\left(y+\frac{w_y}{2}\right)\right] \cdot\\
& \cdot\left[\delta\left(z-\frac{d+s}2\right)+\delta\left(z+\frac{d+s}2\right)\right]
\end{split}
\end{equation}
for the symmetric resonance. Here $\delta(y)$ is the Dirac delta function, $\Theta(y)$ is the Heaviside step function and $q_0$ is the amplitude of the uncompensated charge oscillations at third harmonic frequency. The latter depends on the magnitude of the nonlinear polarization and is proportional to the $\hat{\chi}^{(3)}$ components and the local field factors at the third harmonic frequency $L_{3\omega}(\theta)$ and fundamental frequency $L_\omega (\theta)$ cubed. The solution of the potential equation
\begin{eqnarray}\label{eq:phi}
\label{eq:A}
\left(\Delta-\frac1{c^2}\frac{\partial^2}{\partial t^2}\right){\mathbf A}({\mathbf r},t)=-\mu_0{\mathbf j}({\mathbf r},t)
\end{eqnarray}
is sought. The problem is considered to be two dimensional, i.e., $x$ independent. First, we consider the antisymmetric resonance. The solution of Eq.\ (\ref{eq:A}) could be expressed with the retarded potential
\begin{eqnarray}
\label{eq:Asol}
{\mathbf A}({\mathbf r},t)=\frac{\mu_0}{4\pi}\int dV^\prime \frac{{\mathbf j}({\mathbf r^\prime},t-|{\mathbf r^\prime}-{\mathbf r}|/c)}{|{\mathbf r^\prime}-{\mathbf r}|}.
\end{eqnarray}
Since ${\mathbf H}=\mbox{curl}{\mathbf A}/\mu_0$, the magnetic field distribution in the far field ($r\gg r^{\prime}$) is expressed in the cylindrical coordinates by substitution of Eq.\ (\ref{eq:jas}) into Eq.\ (\ref{eq:Asol}) as follows:
\begin{equation}
\begin{split}
H_x &= -\frac{3\omega q_0\sin\beta}{\pi r \cos\beta}\cdot\sin\left(\frac{k w_y\cos\beta}{2}\right) \cdot \\
& \cdot\sin\left(\frac{k(d+s)\sin\beta}2\right)\cdot\sin(3\omega t-kr),\\
H_y &=H_z=0,\label{eq:Hx}
\end{split}
\end{equation}
where $k=3\omega/c$ and $\beta=\theta+\pi/2$. The angular radiation pattern $R(\theta)$ is defined by the averaged electromagnetic intensity which the unit cell of the metamaterial emits per unit solid angle as a function of radiation angle. It is expressed as follows:
\begin{equation}\label{eq:dIdt}
R(\beta)=\frac{dP}{d\beta}=r\overline{\left[{\mathbf r}\cdot[{\mathbf E}\times{\mathbf H}]\right]}.
\end{equation}
For a plane wave it applies ${\mathbf r}\cdot{\mathbf E}\times{\mathbf H}=r\sqrt{\mu_0/\epsilon_0}H^2$. By substitution of Eq.\ (\ref{eq:Hx}) into Eq.\ (\ref{eq:dIdt}) and time averaging  we get the angular radiation pattern for the antisymmetric resonance:
\begin{equation}\label{eq:ras}
\begin{split}
R_{as}(\beta) & \propto \bigg[ q_0\tan\beta\sin\left(\frac{kw_y\cos\beta}{2}\right) \cdot \\
& \cdot \sin \left(\frac{k(d+s)\sin\beta}2\right) \bigg] ^2.
\end{split}
\end{equation}
The radiation pattern can be evaluated for the symmetric resonance in the same way by use of Eqs.\ (\ref{eq:js}) and (\ref{eq:Asol}):
\begin{equation}\label{eq:rs}
\begin{split}
R_{s}(\beta) & \propto \bigg[ q_0\tan\beta\sin\left(\frac{kw_y\cos\beta}{2}\right) \cdot \\
& \cdot \cos\left(\frac{k(d+s)\sin\beta}2\right)\bigg]^2,
\end{split}
\end{equation}
The polar plots in Fig.\ \ref{fig:fishnet_sketch} show the normalized angular dependences of THG calculated using Eqs.\ (\ref{eq:ras}) and (\ref{eq:rs}) for the antisymmetric and symmetric resonances, respectively, for the same parameters. The dependence of $q_0(\theta)\sim |L_\omega(\theta)|^3$ can be expressed for the magnetic resonance with a Lorentz spectral line:
\begin{equation}\label{eq:q}
L_\omega(\theta)\sim \left[\left(\omega^0_0 +\frac{\partial\omega_0}{\partial\theta}\theta \right)^2 - \omega^2 + 2 i \gamma \omega\right]^{-1}.
\end{equation}
relying on the approximation under which the local field correction factor is proportional to the absorption contour function. Then, the central frequency of the resonance $\omega_0(\theta)$ is substituted by the truncated Taylor expansion in the form of $\omega_0(\theta) = \omega_0^0 + \theta \partial\omega_0/\partial\theta$ accounting for angular dispersion of the resonance. The angular radiation pattern of the third harmonics is straightforwardly connected to the angular dependence of THG. The third harmonic radiation is emitted from each unit cell of the metamaterial with the relative phase depending on the angle of incidence of the pump. Radiation from each cell interferes to compose the diffraction pattern. The intensity of each diffraction lobe depends on the angle of diffraction via the radiation pattern dependence. If only the zeroth diffraction order is detected then the diffraction angle equals the angle of incidence and thus the radiation pattern is probed by measuring the angular dependence of THG. Now we use Eqs.\ (\ref{eq:ras}) and (\ref{eq:q}) to calculate the data on angular-dependent THG from the fishnet metamaterial. The function used is expressed as follows:
\begin{equation}\label{eq:fit}
\begin{split}
I(\theta)&= B\bigg[|L(\theta)|^3 \cot(\theta)\sin\left( \frac{kw_y\sin\theta}{2}\right)\cdot\\
&\cdot\sin\left(\frac{k(d+s)\cos\theta}2\right) \bigg]^2.
\end{split}
\end{equation}
The parameters in Eq.\ (\ref{eq:q}) are determined from the linear measurements, [see Fig.\ \ref{fig:linear}]. The angular dispersion of the resonance central frequency is $\partial\omega_0/\partial\theta\simeq3\cdot10^{12}\,\text{rad}/(\text{deg}\cdot \text{s})$ and $\gamma=0.15\pm0.01$\,ps$^{-1}$ (corresponds to $\Delta \lambda_{\mbox{\tiny FWHM}}=220$\,nm). The parameter $B$ stands for a calibration coefficient that was not measured precisely. For $w_y$ the SEM-measured value was taken and $(d+s)$ was set to $250\,$nm. The angular dependent third harmonic intensity function [Eq.\ \ref{eq:fit}] is plotted in Figs.\ \ref{fig:thg}(a)--4(d) with solid lines. A good quantitative correspondence is observed between the experimental data, the numerically calculated data and the modeled dependence. From all the parameters only $(d+s)$ differs from the experimentally measured one. The main reason is general oversimplification of the model, i.e., not taking the real phase velocity of the third harmonic radiation inside the metamaterial into account, considering pure symmetric or antisymmetric modes, assuming infinitely dense charge and current distributions, etc. Nevertheless, the model gives an explicit way how one can distinguish between symmetric and antisymmetric resonances of the metamaterial by means of its nonlinear optical response. For the symmetric resonance no local extremum is observed at oblique incidence whereas the maximum is present in the case of the antisymmetric resonance. In terms of effective $\hat{\chi}^{(3)}$ tensor components of the metamaterial this means that the ${\chi}^{(3)}_{yyyy}$ component of the medium at the magnetic resonance is less pronounced than that at the electric resonance. In correspondence with the general concept of metamaterials it makes possible to tailor the relation between different tensor components by the proper choice of the metamaterial resonance and its parameters. Moreover, it could be seen from Eqs.\ (9) and (10) that effective nonlinearities of the metamaterial straightforwardly depend on its dimensions, namely $w_y$, $d$, and $s$ in the framework of the model.

\section{Conclusions}

To conclude, a magnetic resonance contribution to third-order optical nonlinearities of the fishnet metamaterial was shown. It was achieved by means of measurements of the third harmonic signal in the forward direction from a fishnet sample and numerical simulations with a nonlinear FMM. Interference of radiation from separated third harmonic sources is shown to emerge as a local maximum in the angular spectra of the third harmonic signal found at oblique incidence. Antisymmetric oscillations of currents, which are the intrinsic properties of magnetic resonances, are found to be responsible for the particular radiation pattern. Based on this an analytical model was built. The angular characteristic of the third harmonic response from the experiment, the FMM, and the analytical model were compared. A quantitative correspondence between these data sets is observed. The results contribute to a better understanding of the possibilities of the nonlinear properties of optical metamaterials with plasmonic resonances of different symmetries.

\section{Acknowledgments}

The authors acknowledge support from the German Research Foundation (SPP 1391 prority program, NanoGuide), the German Federal Ministry of Education and Research (PhoNa, Metamat), the Russian Foundataion for Basic Research, and the Ministry of Education and Science of the Russian Federation.

\section{Appendix A}

Here we discuss the phase difference between the sources of third harmonic radiation. The sources of the radiation are oscillations in the gold layers at the third harmonic frequency. We use a coupled oscillator model with a nonlinear extension. Uncompensated charges are induced at the edges of the thick wires of the metamaterial by the external electromagnetic field with a polarization along the thin wires as shown in Fig.\ \ref{fig:fishnet_sketch}.\cite{Mary2008} Charge conservation implies $q_1(t)=-q_3(t)$ and $q_2(t)=-q_4(t)$. Harmonic oscillations of the charge densities in two coupled layers can be described as a superposition of two eigenmodes of the system\,---\,the first one corresponds to codirectional currents in the layers and the second one corresponds to counterdirectional ones.\cite{Petschulat2008} Consider $x_1(t)=q_1(t)-q_3(t)=2q_1(t)$ for the uncompensated charge at the upper fishnet layer and $x_2(t)=q_2(t)-q_4(t)=2q_2(t)$ for the lower fishnet layer. The linear dynamics of these values is described by the coupled harmonic oscillator model:
\begin{eqnarray}\label{eq:x1}
\ddot{x}_1(t)+2\gamma \dot{x}_1(t)+\omega_0^2x_1(t)+\sigma x_2(t) & = & fe^{i \omega t}\\
\ddot{x}_2(t)+2\gamma \dot{x}_2(t)+\omega_0^2x_2(t)+\sigma x_1(t) & = & fe^{i (\omega t+\varphi_0)}.
\label{eq:x2}
\end{eqnarray}
Here is $\gamma$ the damping constant, $\omega_0$ is the central frequency of the resonance for an isolated layer, $\sigma$ is the coupling constant, $f$ is the oscillator strength and $\varphi_0$ is the difference of phases of the exciting fields caused by the retardation. The dynamics of the asymmetric mode $X(t)=x_1(t)-x_2(t)$ is described by
\begin{equation}
\ddot{X}(t)+2\gamma \dot{X}(t)+\omega_0^2X(t)-\sigma X(t)=f(1-e^{i\varphi_0})e^{i \omega t}.
\end{equation}
The solution of the equation in the frequency domain is expressed as
\begin{equation}
X(\omega)=\frac{f(1-e^{i\varphi_0})}{\omega^2_0-\omega^2+2i \gamma \omega-\sigma}.
\end{equation}
In the case when the $Q$ factor of the modes is high enough for the condition $\sqrt{\sigma}\gg\gamma$ to be held the asymmetric mode implies $x_1(\omega)+x_2(\omega)\approx0$ and $\arg x_1(\omega)-\arg x_2(\omega)=\pi$ as a consequence. 

Now we consider a nonlinear addition to the electron movements
\begin{equation}
\ddot{x}_1(t)+2\gamma \dot{x}_1(t)+\omega_0^2x_1(t)+\sigma x_2(t)+\alpha x_1^3(t)
=fe^{i \omega t}
\end{equation}
and
\begin{equation}\ddot{x}_2(t)+2\gamma \dot{x}_2(t)+\omega_0^2x_2(t)+\sigma x_1(t)+\alpha x_2^3(t)
=fe^{i ( \omega t+\varphi_0)},
\end{equation}
where $\alpha\ll\gamma^2\omega_0^4/f^2$. This restriction corresponds to the experimentally observed low conversion ($\approx10^{-11}$) from the fundamental field to the third harmonic field and allows one to use the perturbation theory approach. At the magnetic resonance apply $x_1(t)\approx-x_2(t)$ and only one equation has to be considered:
\begin{equation}
\label{eq:xpp}
\ddot{x}_1(t)+2\gamma \dot{x}_1(t)+\omega_0^2x_1(t)-\sigma x_1(t)+\alpha x_1^3(t)=fe^{i \omega t}.
\end{equation}
The approximate solution is reduced to two terms:
\begin{equation}
x_1(t)=x_1^0(\omega)e^{i\omega t}+x_1^\prime(\omega) e^{i 3\omega t}.
\end{equation}
After substituting the solution into Eq.\ (\ref{eq:xpp}) and calculating the multipliers of $e^{i\omega t}$ and $e^{i3\omega t}$, one gets
\begin{equation}
x_1^0(\omega)=\frac{f}{\omega^2_0-\omega^2+2i \gamma \omega-\sigma}
\end{equation}
and
\begin{equation}\label{1}
x_1^\prime(\omega)=\frac{\alpha}{\omega_0^2-(3\omega)^2+6i\gamma\omega-\sigma}\left(x_1^0(\omega)\right)^3.
\end{equation}
Analogously one gets
\begin{equation}\label{eq:x1prime}
x_2^0(\omega)=\frac{fe^{i\varphi_0}}{\omega^2_0-\omega^2+2i \gamma \omega-\sigma}
\end{equation}
and
\begin{equation}\label{eq:x2prime}
x_2^\prime(\omega)=\frac{\alpha}{\omega_0^2-(3\omega)^2+6i\gamma\omega-\sigma}\left(x_2^0(\omega)\right)^3.
\end{equation}

Since the first multipliers in Eqs.\ (A10) and (A12) are not resonant and have the same phase, the phase difference $\arg x^\prime_1-\arg x^\prime_2$ is defined by the second multipliers. These multipliers are equal to $\left(x^0_1(\omega)\right)^3$ and $\left(x^0_2(\omega)\right)^3$ for the upper and lower layers, respectively. As a consequence $\arg x^\prime_1-\arg x^\prime_2=3(\arg x_1(\omega)-\arg x_2(\omega))=3\pi$ which means that at the THG frequency the electrons move  inside two gold layers out of phase.

\end{document}